\documentclass[aip,amssymb,amsmath,superscriptaddress,preprint,showkeys]{revtex4-1}

\usepackage{graphicx}
\usepackage{amssymb}
\usepackage{amsmath}
\usepackage{textcomp}
\usepackage{here}

\begin{document}

\title{Field-Effect Devices Utilizing LaAlO$_3$-SrTiO$_3$ Interfaces}

\author{B.~F\"org}
\affiliation{Center for Electronic Correlations and Magnetism, University of Augsburg, 86135 Augsburg, Germany}

\author{C.~Richter}
\affiliation{Center for Electronic Correlations and Magnetism, University of Augsburg, 86135 Augsburg, Germany}
\affiliation{Max Planck Institute for Solid State Research, Heisenbergstra\ss e 1, 70569 Stuttgart, Germany}

\author{J.~Mannhart}
\affiliation{Max Planck Institute for Solid State Research, Heisenbergstra\ss e 1, 70569 Stuttgart, Germany}

\date{\today}

\begin{abstract}

Using LaAlO$_3$-SrTiO$_3$ bilayers, we have fabricated field-effect devices that utilize the two-dimensional electron liquid generated at the bilayers' {\textit n}-type interfaces as drain-source channels and the LaAlO$_3$ layers as gate dielectrics. With gate voltages well below 1\,V, the devices are characterized by voltage gain and current gain. The devices were operated at temperatures up to 100\,$^\circ$C.

\end{abstract}

\maketitle

Silicon field-effect transistors (FETs) are the backbone of modern electronics. They are three-terminal devices with current and voltage gain and are characterized by a vanishing back-action from the output to the input, to name two of their key features. In recent years, complex oxides have become materials of interest for the drain-source (DS) channels of FETs\cite{AHNFET}.  All-oxide FETs are explored to enhance the functionality of FETs and to investigate possible roads to overcome the scaling limitations of silicon-based devices.
Among possible oxide DS channels, the metallic layers generated at oxide interfaces \cite{Highmobility, MHA_Schlom} form a special category, as they are thin, two-dimensional, and possibly correlated electron systems with a low carrier density and typically high low-temperature mobilities (for LaAlO$_3$-SrTiO$_3$ interfaces 10\,cm$^2$/Vs at 300\,K and up to 50,000\,cm$^2$/Vs at 2\,K)\cite{oscillations,Huijben}. Indeed,  early on, large electric field effects \cite{FieldThiel} were found in the guinea pig of oxide 2D interfaces, the {\textit n}-type LaAlO$_3$-SrTiO$_3$ interface \cite{Highmobility}. Owing to the 1-mm-thick SrTiO$_3$ substrates used as gate dielectrics in those studies, turn-on voltages of $V_{\text{G}}= 60$\,V were needed to switch the devices. In another approach, nanometer-sized lateral field-effect devices were fabricated by using the tip of a scanning force microscope to write conducting lines into LaAlO$_3$-SrTiO$_3$ interfaces \cite{Levy}. In these studies, much smaller gate voltages, a few volts, were found to be sufficient to switch the interfaces.
The large band gap (5.6\,eV) and the high dielectric constant of the LaAlO$_3$ films ($\epsilon_r\sim$18)\cite{Edge} make it possible to use the LaAlO$_3$ layers directly as a gate dielectric in a planar geometry. A cross-section of a device is sketched in Fig.\,1. In these devices, due to the LaAlO$_3$-SrTiO$_3$ polar discontinuity, the LaAlO$_3$ gate dielectric  induces a phase transition in the DS-channel, turning it from an insulator into a metal. The YBa$_2$Cu$_3$O$_7$ used as gate contact reduces the channel's carrier density by adding an effective built-in voltage to the gate. The applied gate voltage $V_{\text{G}}$  is used as is usually done to change the channel conductivity by changing the channel's carrier density, allowing the depletion of the channel into the insulating phase. Following this idea we have fabricated field-effect devices that operate at 300\,K with $V_{\text{G}}<1$\,V and show both voltage and current gain.

The samples were fabricated via reflective high-energy electron diffraction controlled pulsed laser deposition to grow 9-unit-cell-thick LaAlO$_3$ layers on Ti-terminated SrTiO$_3$ substrates (at 780\,$^\circ$C and $9\times10^{-5}\,$mbar of O$_2$) using a single crystalline  LaAlO$_3$ target. Subsequently, as gate electrodes 40\,nm of YBa$_2$Cu$_3$O$_7$ were grown in situ at 760\,$^\circ$C and 0.11\,mbar O$_2$. The in situ growth of YBa$_2$Cu$_3$O$_7$ was chosen to lower the density of the interface states at the LaAlO$_3$ surface. The samples were cooled in 400\,mbar of O$_2$ with an annealing step of 1 hour at 600\,$^\circ$C and two 30 minutes steps at 460\,$^\circ$C and 430\,$^\circ$C. The  LaAlO$_3$  was photolithographically patterned using the technique described in Ref.\,[9], the YBa$_2$Cu$_3$O$_7$ by photolithography and etching in 0.5\%\,H$_3$PO$_4$. Electrical contacts to the interface were provided by Ar-ion milled holes refilled in situ with sputtered titanium; gold contacts to the gates were deposited by ex situ sputtering. The layout of the sample and a micrograph are shown in Fig.\,1. The channel width was 1600\,\textmu m, their lengths varied between 200\,\textmu m and 20\,\textmu m. Four samples were fabricated in this manner, and their properties were found to be reproducible and stable for durations of several months. All measurements were performed in darkness to avoid photo-excited carriers. As a precaution against reducing the SrTiO$_3$, measurements above room temperature were performed in 400\,mbar of O$_2$; at all other temperatures, measurements were performed in the He atmosphere of a cryostat. For all measurements presented here, the gate leakage current $I_{\text{G}} < 150\,$nA was small on the scale of the drain-source currents $I_{\text{DS}}$.

Figs.\,2 and 3 show the $I_{\text{DS}}$($V_{\text{GS}}$) dependencies of two devices. The DS channels are self-conducting at all temperatures investigated (Fig.\,3). As expected for {\textit n}-type channels, positive (negative) voltages applied to the gate ($V_{\text{GS}} > 0$) increase (decrease) $I_{\text{DS}}$ (Figs.\,2-4), a change of $V_{\text{G}}$ by $700\,$mV causing a change of $I_{\text{DS}}$ by 4 orders of magnitude. The ratio $R (V_{\text{GS}} = -330\,\text{mV}) / R (V_{\text{GS}} =180\,\text{mV}$), where $R$ is the DS resistance at $V_{\text{DS}}= 0$, exceeds 150 at room-temperature. With a dielectric constant $\epsilon_r=18$ of LaAlO$_3$  films\cite{Edge}, a gate voltage of 300\,mV is estimated to change the channel carrier density by $~10^{12}$/cm$^{2}$. We therefore calculate that, at room temperature and $V_{\text{GS}} =0$, the density of mobile charge carriers equals $\sim10^{12}$/cm$^{2}$, which suggests that the presence of YBa$_2$Cu$_3$O$_7$ depletes the LaAlO$_3$-SrTiO$_3$ interface, in agreement with a previous report \cite{dielectric1}. With decreasing temperature, the characteristics display an enhancement of the conductivity at positive $V_{\text{GS}}$ and a reduction of the turn-on voltage, reflecting the increase of the interface conductance and a reduction of the mobile carrier density with cooling.  The saturation mode was not reached by the $V_{\text{DS}}$ applied in these studies.

Figure 5 shows for a set of $I_{\text{DS}}$ values the dependence of $V_{\text{DS}}$ on $V_{\text{GS}}$, again at 300 K. The $G=\Delta V_{\text{DS}} / \Delta V_{\text{GS}} (V_{\text{GS}})$ dependence is displayed in the inset. As shown, a voltage gain~$G>1$ is readily obtained, with $G\sim 40$, for $I_{\text{GS}}=5\,$\textmu A and $V_{\text{DS}}=450\,$mV, to give an example.

In summary, using LaAlO$_3$-SrTiO$_3$ bilayers we have fabricated all-oxide field-effect transistors with current and voltage gain. The FETs were operated up to 100\,$^\circ$C. Our studies show that the transport properties of a 2-D electron system generated at an oxide interface can be effectively controlled by using gate fields induced by a top-gate configuration. Because then the gate dielectric is only a few monolayers thick, the devices can be operated with small gate voltages, the lower limits of which remain to be explored. Optimization of DS-channel and of the gate stack are suggested to further enhance the performance of such oxide field effect devices. Here we note, that an ultra-thin gate stack comprising two unit cells of LaAlO$_3$ and two unit cells of SrTiO$_3$ already generates a 2-D electron system\cite{LAO-STO-stack}.

The authors gratefully acknowledge helpful discussions with W.~Haensch, R.~Jany, C.~Hughes, B.~Keimer, and H.~Riehl, as well as financial support from the German Science Foundation (TRR 80).

\newpage
\section*{Captures}

FIG.\,1. Sketch of a cross section of a device (a) and electron microscope image of a typical sample (b). The colors were added. The horizontal lines within the LaAlO$_3$-layer (LAO) symbolize the 9 monolayers of LaAlO$_3$, the standard thickness of gate dielectric in this study, grown on a SrTiO$_3$ substrate (STO). The narrow, straight line (red) denotes the location of the cross section shown in (a); the numbers indicate the gate widths in microns. The two-dimensional electron liquid shown in (b) in pale gray is also present under the YBa$_2$Cu$_3$O$_7$ (YBCO) gates (dark gray).

FIG.\,2. Gate-voltage ($V_{\text{GS}}$)-dependent $I_{\text{DS}}$($V_{\text{DS}}$) characteristics of a device (channel length 60\,\textmu m, channel width 1600\,\textmu m) measured in four-point configuration at room temperature.

FIG.\,3. $I_{\text{DS}}(V_{\text{DS}})$-characteristics of a device measured in four-point configuration at $-100$, $20$, and $100$\,$^{\circ}$C. The measurement was done on a device with channel length of 40\,\textmu m and channel width of 1600\,\textmu m.

FIG.\,4. Drain-source current $I_{\text{DS}}$ measured as a function of gate voltage $V_{\text{GS}}$. The inset shows the gate current $I_{\text{GS}}$ as a function of gate voltage $V_{\text{GS}}$ for $I_{\text{DS}}=0$. The data were taken at room temperature on the device shown in Figs.\,2 and 5 (channel length 60\,\textmu m, channel width 1600\,\textmu m).

FIG.\,5. $V_{\text{DS}}$ measured as a function of applied gate voltage $V_{\text{GS}}$ at room temperature. The measurement was done on the device shown in Figs.\,2 and 4 (channel length 60\,\textmu m, channel width 1600\,\textmu m). The inset shows $G=dV_{\text{DS}}/dV_{\text{GS}}$ calculated as derivative on a logarithmic scale.

\clearpage
\begin{figure}
\begin{centering}
\includegraphics{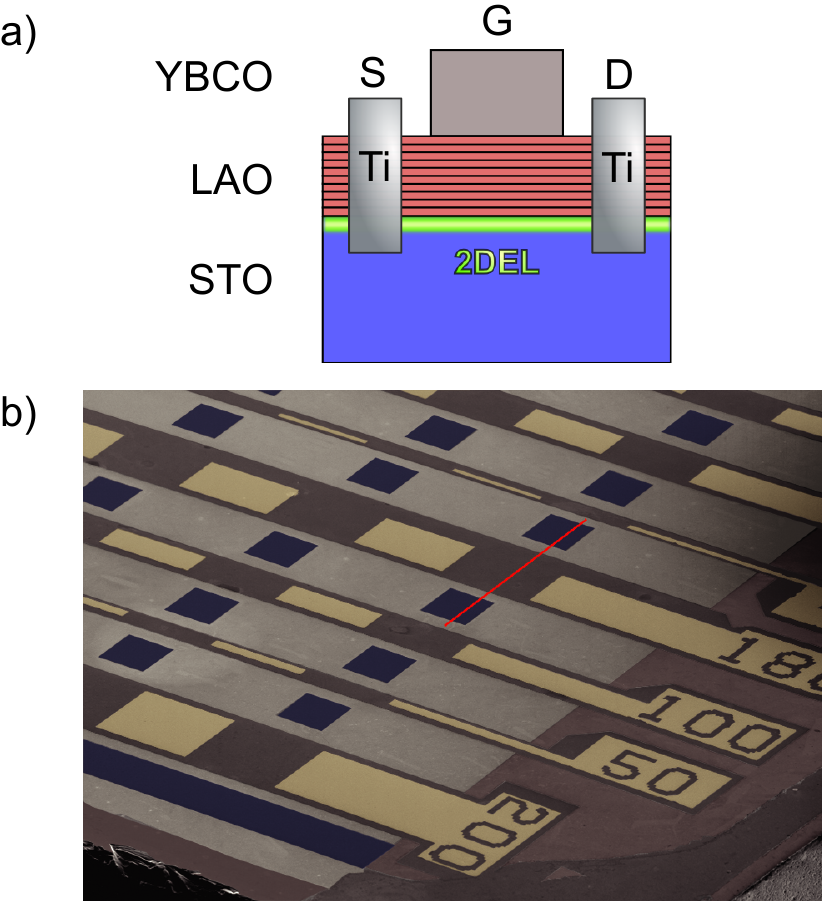}
\end{centering}
\end{figure}

\clearpage

\begin{figure}
\begin{centering}
\includegraphics{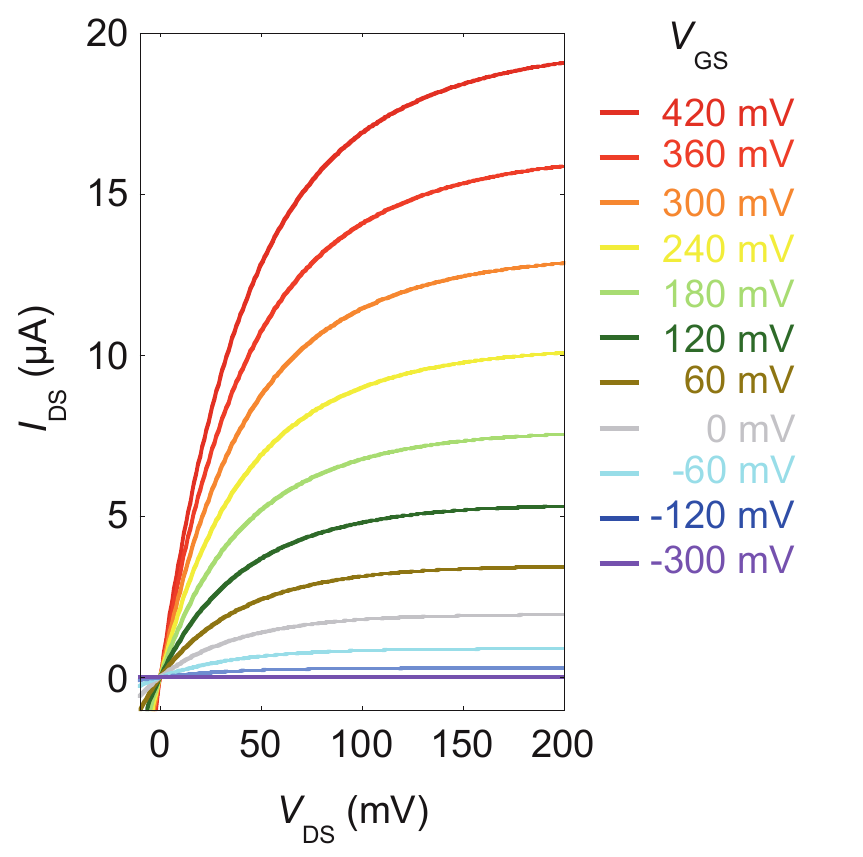}
\end{centering}
\end{figure}

\clearpage

\begin{figure}
\begin{centering}
\includegraphics{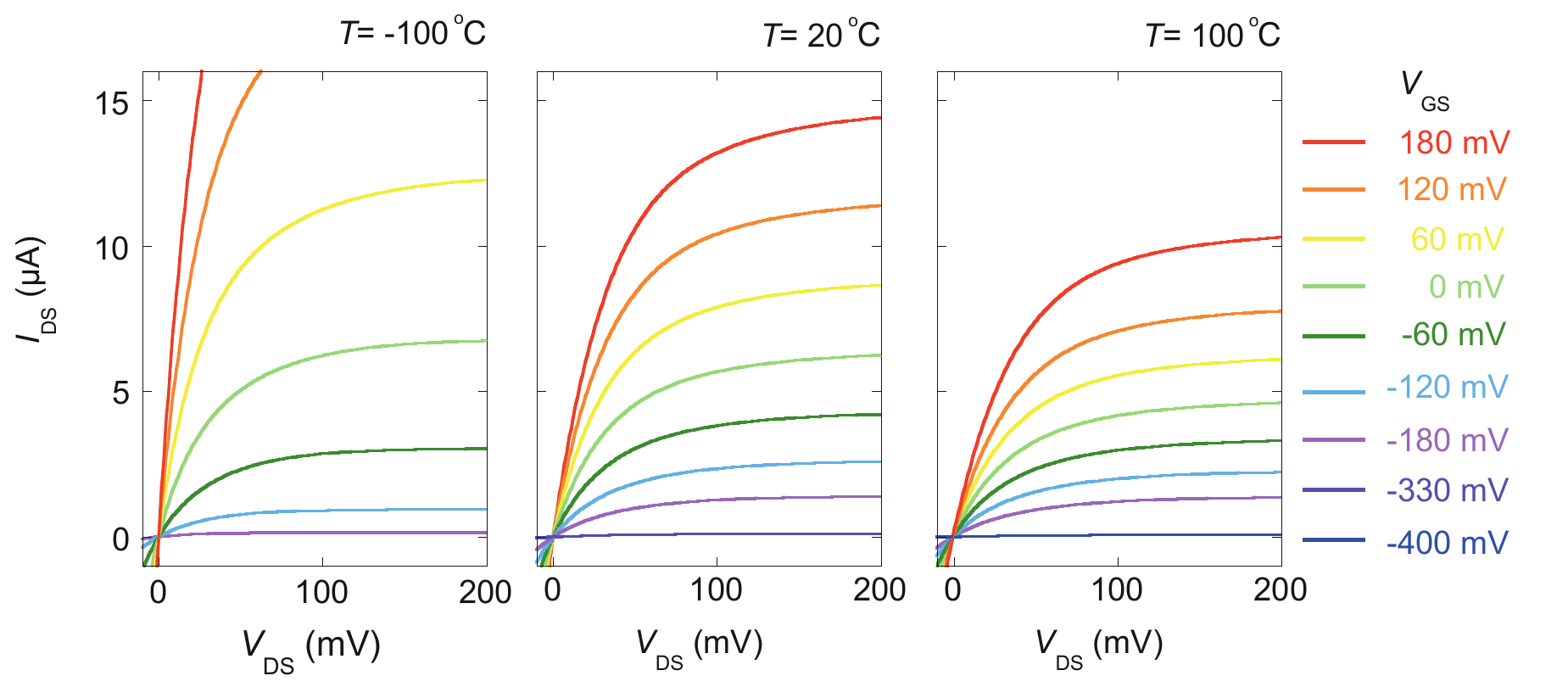}
\end{centering}
\end{figure}

\clearpage

\begin{figure}
\begin{centering}
\includegraphics{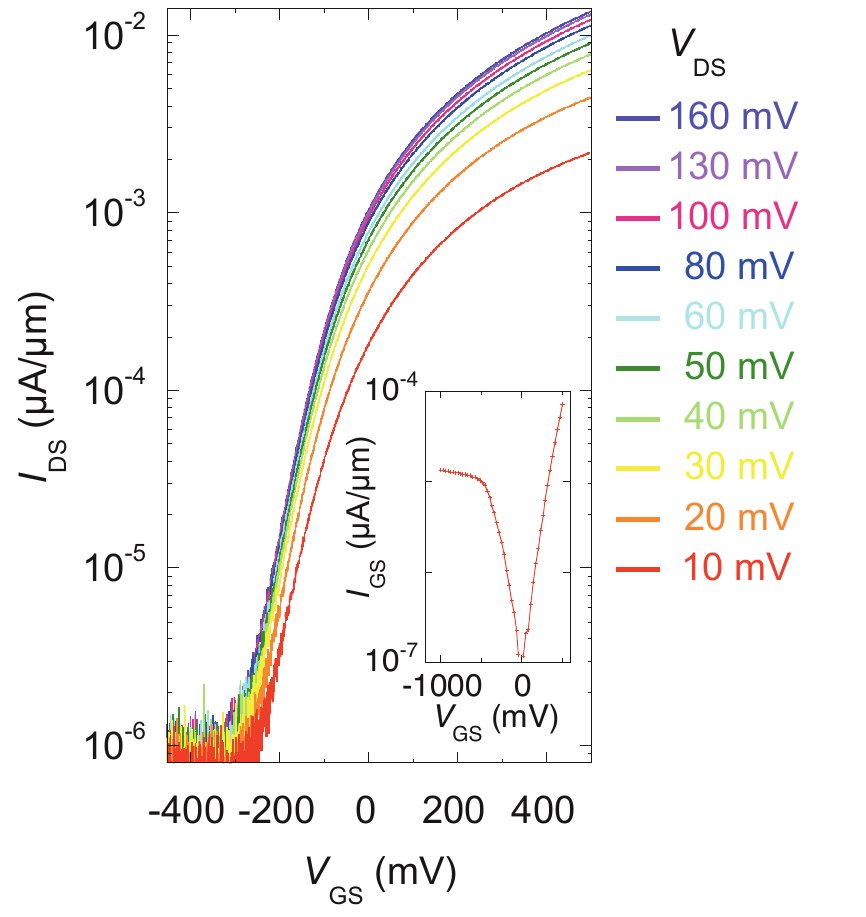}
\end{centering}
\end{figure}

\clearpage

\begin{figure}
\begin{centering}
\includegraphics{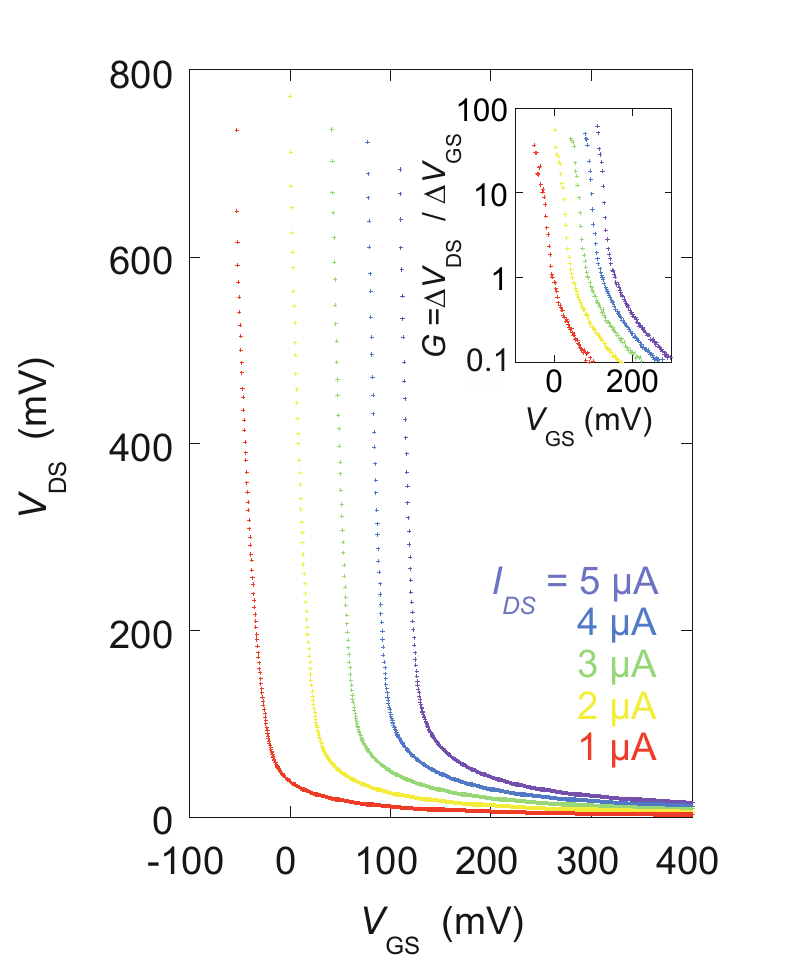}
\end{centering}
\end{figure}

\end{document}